\documentstyle[prl,aps]{revtex}
\begin{document}
\draft
\title{
Initial stages of thin film growth in the presence of island-edge barriers}
\author{Daniel Kandel\cite{email}}
\address{Department of Physics of Complex Systems,\\
Weizmann Institute of Science, Rehovot 76100, Israel}
\maketitle

\begin{abstract}
A model of submonolayer thin film growth is studied, where the attachment
of atoms to island edges is hindered by an energy barrier. A novel
behavior of the density of islands, $N_s$, is predicted as a function
of flux $F$ and temperature $T$. For example,
$N_s$ scales as $F^X$ with $X=2i^{\ast}/(i^{\ast}+3)$, 
where $i^{\ast}$ is the critical
island size, in contrast with the standard result $X=i^{\ast}/(i^{\ast}+2)$.
The theory is applicable to surfactant mediated
growth and chemical vapor deposition. 
It explains recent experiments, which are inconsistent with 
the standard theory.
\end{abstract}


\setlength{\baselineskip}{12pt}
Thin film growth processes are of tremendous importance for the fabrication
of nanostructures and electronic devices. Technological advances in
device miniaturization depend largely on the degree to which one can
control the growth process, the epitaxial quality of the film and the
morphology of the surface. It is therefore essential to understand the
microscopic processes involved in thin film growth and especially their
effect on the structure of the film. 

Of particular interest are the initial stages of growth or the 
submonolayer regime, which is relatively easy to investigate
both experimentally and theoretically. 
It is possible to learn from such studies
about the relevant microscopic processes and their respective energy 
barriers. For example, when the film evolves by nucleation, growth and
coalescence of two dimensional (2D) islands, the shape and size distributions
of the islands and their dependence on physical parameters such as
temperature, flux and coverage, yield information about various
diffusion processes, attachment and detachment of atoms to and from 
island edges, etc. \cite{venables}.

The first theories of diffusion of atoms in the presence of
steps assumed that step edges are perfect sinks for adatoms \cite{bcf}.
This assumption was later relaxed \cite{chernov}, and kinetic coefficients
were introduced to take into account the finite energy barriers 
\cite{ehrlich} associated with attachment and
detachment of adatoms to and from the edges.
The perfect sink assumption has been shown to yield reasonable results
under many experimental conditions.
It was realized, however, that it fails in several
important cases. 
For example, a significant barrier for attachment of adatoms to steps
from the terrace above, leads to a kinetic instability of the
flat surface and to the growth of large mounds \cite{mound}. 
An asymmetry in the
barriers for adatom attachment from below and above the step can also
lead to peculiar step bunching \cite{bunch}
as well as fingering \cite{finger} instabilities.
Island-edge barriers may be important in surfactant mediated growth,
where a surfactant atom can bind to an island edge. In order for an
adatom to attach to the edge, surfactant atoms have to be removed
from there.
Kandel and Kaxiras explained \cite{kankax,tim1}
experimental results related to surfactant
mediated growth by assuming that
the energy barrier for such a change in the atomic configuration is
large \cite{markov}. 
Island-edge barriers may also occur in chemical vapor
deposition (CVD). For example, during CVD of Si on Si using disilane, the 
surface is covered with hydrogen, which can bind to island
edges similarly to surfactant atoms \cite{tim2}.

In this work, the effect of island-edge barriers on submonolayer growth
is studied in the framework of rate equation theory
\cite{venables,stowell,bales}. In particular,
the density of 2D islands on the surface, $N_s(F,T,\theta)$, 
is calculated as a function
of flux $F$, temperature $T$ and coverage $\theta$.
It is shown that island-edge barriers have a dramatic,
experimentally observable, effect
on the behavior of $N_s$. Although finite island-edge barriers have
been studied before \cite{others,bz}
in the context of submonolayer growth, it is the first time their
detailed effect on the dependence of the island density 
on flux and temperature is addressed.

The simplest scheme for the calculation of $N_s$ is the critical island
approximation, where it is assumed that islands that contain more
than $i^\ast$ atoms are stable, while smaller ones are not and can
decay \cite{criti}. 
Within this scheme, $N_s$ is the density of {\em stable} islands,
and a detailed balance relation is assumed to hold 
\cite{walton} between the densities
of unstable islands, $N_i$ ($i\le i^\ast$), and the average adatom
density $\bar{n}$:
\begin{equation}
\Omega N_i=(\Omega\bar{n})^i e^{\beta E_i}~~,
\label{omni}
\end{equation}
where $\Omega$ is the atomic area of the solid, 
$\beta=1/k_B T$ and $E_i$
is the binding energy of an island of $i$ atoms.

Now one can write down the rate equation for the density of stable islands:
\begin{equation}
\frac{dN_s}{dt}=I-C~~,
\label{rate}
\end{equation}
where $C$ and $I$ are the coalescence and nucleation rates per unit area.
When the coverage is small,
coalescence does not occur and $C\approx 0$. Since this is
the limit of interest in this work, $C$ is completely neglected
below.
The nucleation
rate is
\begin{equation}
I=\sigma \bar{n} N^\ast S^\ast~~.
\label{nrate}
\end{equation}
Here $\sigma$ is the capture coefficient of a critical island, 
$N^\ast=N_{i^\ast}$
and
\begin{equation}
S^\ast=\nu\Omega e^{-\beta(E_d+E_b^\ast)}~~.
\label{sstar}
\end{equation}
In the equation above, $E_d$ is the diffusion barrier and $E_b^\ast$ is the
additional barrier for attachment of adatoms to the edge of a critical island
(the limit where the island edge is a perfect sink is obtained by 
taking $E_b^\ast=0$).
$\nu$ is the attempt frequency, assumed to be the same for all
microscopic processes. Thus $S^\ast=D \exp(-\beta E_b^\ast)$, where $D$
is the diffusion constant.

$N^\ast$ can be expressed in terms of $\bar{n}$ via Eq.\
(\ref{omni}), and thus
the nucleation term, $I$, is a function of the average
adatom density.
To estimate $\bar{n}$, consider the density
of adatoms, $n(r)$, 
around a typical stable island of radius $R$
($r$ is the distance from the center of the island and
radial symmetry is assumed).
Under conditions of complete condensation, where no evaporation
occurs, $n(r)$ obeys the diffusion equation
\begin{equation}
D\left[\frac{d^2 n(r)}{dr^2}+\frac{1}{r}\frac{dn(r)}{dr}\right]+F=0~~,
\label{diff}
\end{equation}
where the quasi-static approximation has been used, suppressing the
time derivative of the adatom density. This approximation is valid when
diffusion is fast enough so that at each instance of time, $n(r)$ reaches
a quasi-steady state, where the flux is almost entirely compensated
by the attachment of adatoms to stable islands.
The relevant solution of Eq.\
(\ref{diff}) obeys the following boundary conditions:
\begin{equation}
\begin{array}{rl}
i) & 
\left.\frac{\displaystyle dn}{\displaystyle dr}\right|_{r=R}=
\frac{\displaystyle 1}{\displaystyle \sqrt{\Omega}}\frac{\displaystyle S}
{\displaystyle D-S}n(R)\\
&\\
ii) & 
\left.\frac{\displaystyle dn}{\displaystyle dr}\right|_{r=L}=0~~,
\end{array}
\label{bc}
\end{equation}
where $S=D \exp(-\beta E_b)$, $E_b$ is the additional energy barrier
for attachment of adatoms to the edge of a stable island, and
$L$ is half the distance between stable islands ($N_s=1/\pi L^2$).
The first boundary condition holds for large islands, and was derived by
Bales and Zangwill \cite{bz}.

The solution of these equations is
\begin{equation}
n(r)=\frac{FL^2}{2D}\ln
\frac{r}{R}+
\frac{F}{4D}(R^2-r^2)+
\frac{F\sqrt{\Omega}L^2}{2R}
\frac{D-S}{DS}
\left(1-\frac{R^2}{L^2}\right)~~.
\label{sol}
\end{equation}
Following Stowell and coworkers \cite{stowell}, 
it is easy to obtain an expression for $\bar{n}$:
\begin{equation}
\begin{array}{ll}
\bar{n} & =\frac{\displaystyle 1}{\displaystyle \pi (L^2-R^2)}
{\displaystyle \int_R^L 2\pi r n(r) dr}\\&\\
& \approx
\frac{\displaystyle F}{\displaystyle 4\pi D}(-\ln\theta-
\frac{\displaystyle 3}{\displaystyle 2})
\frac{\displaystyle 1}{\displaystyle N_s}+
\frac{\displaystyle F}{\displaystyle 2S}
\frac{\displaystyle D-S}{\displaystyle D}
\sqrt{\frac{\displaystyle \Omega}{\displaystyle \pi\theta N_s}}~~,
\end{array}
\label{nbar}
\end{equation}
where it is assumed that the density of stable islands is already
large enough so that $\theta\approx R^2/L^2$ and the contribution
of $\bar{n}$ to the coverage is negligible. Since this theory is
valid only in the small coverage limit, terms that vanish when
$\theta\longrightarrow 0$ were omitted.

Eq.\ (\ref{rate}) can now be rewritten as
\begin{equation}
\frac{dN_s}{d\theta}=
\frac{\sigma\Omega^{i^\ast-2}}{F}e^{\beta E^\ast}
S^\ast\bar{n}^{i^\ast+1}~~,
\label{ratea}
\end{equation}
where the relation $\theta=F\Omega t$ was used, $\bar{n}$ is given
by Eq.\ (\ref{nbar}) and $E^\ast=E_{i^\ast}$.

Eq.\ (\ref{ratea}) can be solved numerically
starting from some initial condition $N_s(F,T,\theta_i)=N_s^{(i)}$
($\theta_i\ne 0$ since the quasi-static approximation does not
hold when $\theta=0$). But it is useful to
consider two limiting cases where the problem 
can be solved analytically.
These limits occur when one of the two terms in expression 
(\ref{nbar}) for $\bar{n}$ is small and can be neglected. More
quantitatively, the ratio of the two terms is 
\begin{equation}
G(F,T,\theta)=\frac{2\sqrt{\pi\Omega}}{\sqrt{\theta}(-\ln\theta-3/2)}
(e^{\beta E_b}-1)\sqrt{N_s}~~.
\label{g}
\end{equation}
If $G$ is much smaller than 1 (limit I) for $\theta>\theta_i$, 
the second term on the r.h.s. of Eq.\ (\ref{nbar})
can be neglected. This is 
the perfect sink limit 
$S\longrightarrow D$ ($E_b\longrightarrow 0$), 
and in this case standard results \cite{venables}
are expected to hold. In limit II,
on the other hand, $G\gg 1$
for $\theta>\theta_i$, and 
the second term in (\ref{nbar}) dominates; this is the case where
$\exp(\beta E_b)\gg 1$, and
the island-edge barrier is most important. In limit I, attachment of 
adatoms to island edges is infinitely fast and the kinetics is
diffusion limited, whereas in limit II diffusion is fast, and the kinetics
of adatoms is limited by the slow
attachment to island edges. In both limits, Eq.\ (\ref{ratea}) can
be written in the form
\begin{equation}
\frac{d(N_s\Omega)^\Gamma}{d\theta}
=\Gamma\sigma H(\theta) e^{\beta{\cal{E}}}
\left(\frac{\Omega F}{\nu}\right)^{i^\ast}~~.
\label{lim}
\end{equation}
Taking for simplicity $E_b^\ast=E_b$, one gets in limit I
\begin{equation}
\begin{array}{rcl}
\Gamma=i^\ast+2 & , & {\cal{E}}=E^\ast+i^\ast E_d\\
&\mbox{and}&\\
H(\theta)&=&\left(
\frac{\displaystyle -\ln\theta-3/2}{\displaystyle 4\pi}\right)^{i^\ast+1}~~,
\end{array}
\label{limI}
\end{equation}
while in limit II
\begin{equation}
\begin{array}{rcl}
\Gamma=\frac{\displaystyle i^\ast+3}{\displaystyle 2}&,
&{\cal{E}}=
E^\ast+i^\ast(E_d+E_b)\\
&\mbox{and}&\\
H(\theta)&=&\left(
\frac{\displaystyle 1}{\displaystyle \sqrt{4\pi\theta}}\right)^{i^\ast+1}~~.
\end{array}
\label{limII}
\end{equation}

The solution of Eq.\ (\ref{lim}) is
\begin{equation}
(N_s\Omega)^\Gamma-(N_s^{(i)}\Omega)^\Gamma=\Gamma\sigma e^{\beta{\cal{E}}}
\left(\frac{\Omega F}{\nu}\right)^{i^\ast}
\int_{\theta_i}^
\theta H(\theta')d\theta'
~~.
\label{solim}
\end{equation}
Assuming $N_s^{(i)}\ll N_s(\theta)$, the second term on the l.h.s. of
(\ref{solim}) can be neglected and $N_s$ takes the form
\begin{equation}
N_s(F,T,\theta)=[\Gamma\sigma\int_{\theta_i}^
\theta H(\theta')d\theta']^{1/\Gamma}e^{\beta{\cal{E}}/
\Gamma}\left(\frac{\Omega F}{\nu}\right)^{i^\ast/\Gamma}~~.
\label{solfin}
\end{equation}
Therefore, in limit I
\begin{equation}
N_s\sim e^{\beta(E^\ast+i^\ast E_d)/(i^\ast+2)}~F^{i^\ast/(i^\ast+2)}~~,
\label{solI}
\end{equation}
while in limit II
\begin{equation}
N_s\sim e^{2\beta[E^\ast+i^\ast(E_d+E_b)]/(i^\ast+3)}~
F^{2i^\ast/(i^\ast+3)}~~.
\label{solII}
\end{equation}
The coverage dependence of $N_s$ also differs in the two limits. However,
it is known \cite{rat,bales} that the methods that have been used 
in this work are not suitable for an accurate calculation of this
dependence. For this reason, only the flux and temperature
dependence of $N_s$ are emphasized
in Eqs. (\ref{solI}) and (\ref{solII}).

As expected (see above), in limit I the result 
(Eq.\ (\ref{solI})) coincides with the standard
result calculated under the assumption that island edges are perfect
sinks for adatoms. Eq.\ (\ref{solII}) clearly shows that in limit II,
when island-edge barriers are important, the behavior of $N_s$ as
a function of $F$ and $T$ is strikingly different \cite{sam}. 
The most unambiguous information about the importance of island-edge
barriers in a specific experimental system can be obtained from
the functional dependence of the island density on flux. It is a simple
power law, $N_s\sim F^X$, with an exponent $X$ that depends
only on the critical island size $i^\ast$. While in limit I the
exponent is in the range $1/3\le X_{\text{I}}\le 1$, 
in limit II it can be larger
than 1 ($1/2\le X_{\text{II}}\le 2$). 
For a given value of $i^\ast$, $X_{\text{II}}$
is significantly larger than $X_{\text{I}}$, a difference that can be detected
experimentally. If one has some information about $i^\ast$, a measurement
of $N_s(F)$ can indicate which of the two limits is more appropriate
for the experimental system in question. Furthermore, if there is no
a priori knowledge of $i^\ast$, one can still identify the relevant limit
(and thus evaluate $i^\ast$) if $X<1/2$ or $X>1$. The former case is possible
only in limit I, and the latter only in limit II. Once the proper limit
has been identified and $i^\ast$ evaluated, the temperature dependence
of $N_s$ can be used to estimate 
energy barriers; in particular, one can evaluate ${\cal{E}}$.

Which experimental systems are suitable for such a study?
Homoepitaxial growth experiments of semiconductors and metals
with and without surfactants may be good candidates.
For example, consider the experiments of Voigtl\"{a}nder and Zinner
\cite{voig} on
submonolayer growth of Si/Si(111) with Sb as a surfactant.
In this case, the additional island-edge barrier $E_b$ corresponds to
$E_{ex}-E_d$, where $E_{ex}$ is the barrier for exchange of an adatom with
a surfactant atom at the edge of a stable island, and $E_d$ is the barrier
for diffusion {\em on top} of the surfactant layer.
Limit II corresponds
to the model proposed by Kandel and Kaxiras
\cite{kankax,tim1} to describe surfactant
mediated film growth; they assumed that surfactant atoms passivate island
edges.
Limit I is associated with the more standard
approach to the same problem \cite{surfact}, where no island-edge passivation
is taken into account.
An experimental value of $X>1$ would therefore indicate that the model of 
Kandel and Kaxiras is adequate for this system. A value of $X<1/2$, on the
other hand, would favor the standard approach to surfactant mediated epitaxy.
Intermediate results would be inconclusive.

It is instructive to use experimental numbers in Eq.\ (\ref{g}) for the 
function $G$, and estimate what $E_b$ should be in order for the
system to be in limit II. For example, at $T=900K$ Voigtl\"{a}nder and Zinner
measured an island density of $N_s\approx 10^{11}$ $cm^{-2}$ in the case
of Sb mediated growth (see Fig.\ 3 of Ref.\ \cite{voig}). The coverage was
0.15 bilayers, and since $\Omega\approx 10^{-15}$ $cm^2$, the estimate of
$G$ is $G\approx 0.23(\exp(\beta E_b)-1)$. Therefore, for $E_b=0.4$ eV
$G\approx 39$, and the system should be in limit II. Limit I would be clearly
observed only if $E_b<0.1$ eV. 

Another work of relevance here is the experiment of
Andersohn et al. \cite{andersohn}, 
who measured the exponent $X$ for homoepitaxial growth
of silicon with molecular beam epitaxy (MBE) and CVD using disilane 
(Si$_2$H$_6$).
They concluded that in MBE $X\approx 0.75$, which corresponds to $i^\ast=5-7$
assuming that the standard limit I applies. 
In CVD, on the other hand, they obtained $X\approx 1.25$.
As they emphasize, standard rate equation theory cannot explain
this result, since it always predicts $X<1$. In the framework of the present
work, the result $X=1.25$ is a natural consequence of a significant 
island-edge
barrier (limit II) and corresponds to $i^\ast=5$. In fact, since during 
CVD with disilane, hydrogen is always present on the surface of the 
growing film, it is possible that hydrogen atoms bind to island edges
and have to be displaced before a silicon atom can attach to the island.
In this sense, hydrogen acts as a surfactant in this system \cite{tim2}, 
and the
present interpretation of the experimental result favors the mechanism
of Kandel and Kaxiras
over the standard surfactant mediated growth mechanism
in this particular case.

Here again it is possible to use the experimental numbers in Eq.\ (\ref{g}).
The CVD experiments at $T=800K$ show that for different values of the flux,
the island density at coverage of 0.15 bilayers
changes between $10^{11}$ and $10^{12}$ $cm^{-2}$
(see Fig.\ 4 of Ref.\ \cite{andersohn}). Taking
the smaller of the two values with $E_b=0.4$ eV, one gets $G\approx 75$, and
the system should exhibit a limit II behavior. Thus, the analyses of both
experiments shows that the values of the additional barrier, $E_b$, need
not be rediculously large for the system to be in limit II.

It should be emphasized that the CVD experiments interpreted above are
not a reliable test of the theory due to the complexity of the processes
involved in CVD. Hopefully, this work will stimulate experiments on
simpler systems that will enrich our understanding of the role of
surfactants in thin film growth.

In summary, this paper shows that island-edge barriers have a profound
effect on the density of 2D islands during submonolayer epitaxy, and
as a result on the morphology of the growing surface. The novel behavior
of the island density, predicted in this case, has been shown to be
relevant for experimental systems, including surfactant mediated epitaxy
and chemical vapor deposition. 

I would like to thank E. Kaxiras, 
P. \v{S}milauer and D. Wolf for
stimulating discussions,
B. Voigtl\"{a}nder for drawing my
attention to Ref.\ \cite{andersohn}, and A. Zangwill 
for discussions and for making 
Ref.\ \cite{bz} available prior to publication.
D.K. is the incumbent of the Ruth Epstein Recu
Career Development Chair.


\begin{thebibliography}{10}
\bibitem{sam}
Note that the value of the exponent $X$ in the novel limit II is consistent
  with Eq.\ (4) of Ref.\ \cite{rat} if one assumes $q=0$ and $p=1/2$.
\end{thebibliography}
%

\end{document}